 \documentstyle[12pt]{article}
\begin{document}

\title{Vortex -- Kink Interaction and Capillary Waves in a Vector Superfluid}
\author{ A.A. Nepomnyashchy$^{a,c}$ and L.M. Pismen$^{b,c}$ \\
${}^a$ Department of Mathematics \\
${}^b$ Department of Chemical Engineering \\
${}^c$ Minerva Center for Nonlinear Physics of Complex Systems \\
Technion -- Israel Institute of Technology, 32000 Haifa, Israel}
\maketitle

\begin{abstract}

Interaction of a vortex in a circularly polarized superfluid component of a
2d complex vector field with the phase boundary between superfluid phases
with opposite signs of polarization leads to a resonant excitation of a
``capillary'' wave on the boundary. This leads to energy losses by the
vortex--image pair that has to cause its eventual annihilation.

\end{abstract}

\newpage

\paragraph{Introduction.} Several recent publications
considered dynamics of a 2d complex vector field, that can be
related both to transverse nonlinear optical patterns in polarized light
\cite{gil,lp,max}
and to the motion of a hypothetical fermionic superfluid \cite{pom}.
Although dissipation may be non-negligible in optical systems, radiation
and dispersion effects play there a primary role. In a scalar case,
``acoustic" radiation is the principal mechanism of relaxation to lower
energy states. In the vector field, that can be interpreted as a combination
of two either miscible  or immiscible superfluids, a new relaxation
mechanism may appear: radiation of
capillary waves. In this communication, we explore the influence of this
effect on the interaction of a vortex with an interface between two
immiscible superfuids.

Conservative
Galilean dynamics of a 2d complex vector field is described by an evolution
equation derived from the Hamiltonian $\cal H$ (see \cite{lp}):
  \begin{equation}
  {\bf u}_t=- i\frac{\delta {\cal E}}{\delta {\bf u}^*}, \;\;
{\cal H} = \int [\nabla {\bf u} : \nabla {\bf u}^*  +  V({\bf u})]
d^2 {\bf x} .
    \label{rgle}   \end{equation}
%Here the asterisk denotes the complex conjugate; it is assumed that
%the double vector product does not mix vector and gradient
%components.
The simplest form of a potential that possesses required
symmetries to spatial rotations and phase translations but breaks the
maximal SU(2) symmetry is
\begin{equation}
V({\bf u})= \frac{1}{2} \left[ (1 - {\bf u} \cdot {\bf u}^*)^2 +
   \gamma ({\bf u} \cdot {\bf u} ) ( {\bf u}^* \cdot {\bf u}^*) \right] ,
   \label{rglv}   \end{equation}
%
%The field can be viewed as consisting of two superfluid components
%circularly polarized in the opposite sense \cite{lp}. In this
%representation,
In the representation \cite{lp}
\begin{equation}
 {\bf u} =  u_+ {\bf U}  + u_-  {\bf U}^*; \;\;\;
{\bf U} = \frac{1}{\sqrt{2}} \left| \begin{array}{c}
  \exp ( i \pi/4)\\  \exp ( -i \pi/4) \end{array}\right|
 =\frac{1}{2} \left| \begin{array}{c} 1 + i \\ 1 - i \end{array}\right| ,
\label{basvec} \end{equation}
%and
 the dynamic equations take the form
\begin{equation}
 -i\partial _t u_\pm =\nabla^2 u_\pm +
    \left[1 - |u_\pm|^2 - (1 + 2\gamma) |u_\mp|^2 \right] u_\pm .
 \label{rglfpm} \end{equation}

By setting $u_\pm=\rho_\pm^{1/2}\exp\left(\frac{i}{2}\phi_\pm \right)$,
these equations can be brought to a ``fluid-mechanical'' form:
 \begin{equation}
 \partial_t  \rho_\pm +  \nabla\cdot( \rho_\pm {\bf v}_\pm) = 0,\;\;\;
\partial_t \phi_\pm +  \frac{1}{2} |{\bf v}_\pm|^2  + p_\pm = 0,
  \label{bern} \end{equation}
where
  $\rho_\pm$ are interpreted as ``densities'',
${\bf v}_\pm =
\nabla \theta_\pm$, as ``velocities''
and
%. The dynamic equations obtained after
%separating the real and imaginary parts are the continuity equations
%  \begin{equation}
% \partial_t  \rho_\pm +  \nabla\cdot( \rho_\pm {\bf v}_\pm) = 0
%   \label{cont} \end{equation}
% and the Bernoulli equations
%  \begin{equation}
% \partial_t \theta_\pm +  \frac{1}{2} |{\bf v}_\pm|^2  + p_\pm = 0,
%  \label{bern} \end{equation}
%where
``pressures'' $p_\pm$ obey the ``equations of state''
\begin{equation}
 - p_\pm =  \frac{1}{2\sqrt{ \rho_\pm}}\nabla ^2\sqrt{ \rho_\pm}
      + 1 -  \rho_\pm - (1 + 2\gamma)  \rho_\mp .
 \label{state} \end{equation}

The two superfluids are ``miscible'' at $\gamma<0$, and tend to separate at
$\gamma>0$. In the latter case, two types of topological defects are
possible: vortices in either superfluid and kinks separating two oppositely
polarized superfluid phases.
%Vortices may have either hollow (``punched'')
%core structure, or a repolarized core (i.e. ``filled'' by the oppositely
%polarized component) \cite{lp}. The latter structure is more energeticaly
%favorable at small $|\gamma|$.

The stationary structure of a straight kink is defined by the equations
\begin{equation}
\psi_\pm''  + \psi_\pm[1 -
    \psi_\pm^2 -(1+2\gamma) \psi_\mp^2] =0,
\label{cpr}   \end{equation}
where $\rho^\pm=\psi_\pm^2$, and primes denote derivatives with respect to
the coordinate $y$ directed normally to the kink. The asymptotic conditions
are $\psi_\pm(\pm\infty) = \pm1, \; \psi_\pm(\mp\infty) = 0$. The kinks of
this kind were studied formerly in a context of dissipative systems
\cite{mnt} and dispersive Kerr media \cite{hs}. Recently, the deformation of
kinks has been studied numerically in an optical context \cite{sh}.

Being the only parameter of the model, $\gamma$ defines both the thickness
of the kink and the energy per unit length, or {\em line tension}
$$\sigma=\int_{-\infty}^{\infty}[(\psi_+^{\prime 2}+\psi_-^{\prime 2})+
\frac{1}{2}(\psi_+^2+\psi_-^2-1)^2+2\gamma \psi_+^2\psi_-^2]dy.$$
%At $\gamma \gg 1$, the excess energy of the kink is concentrated in a
%narrow $O(\gamma^{-1/2})$ layer, and $\sigma =O(\gamma^{1/2}) \gg 1$.
%Similar scaling considerations show that $\sigma =O(\gamma^{1/2})$ also at
%$\gamma \ll 1$. The energy and the characteristic core size of vortices are
%defined by the parameter $\gamma$ in a different way: the size is of
%$O(\gamma^{-1/2})$ only at $\gamma \ll 1$, while at $\gamma \le 1$ it is of
%$O(1)$. The energy is scaled as $\ln(R\gamma^{1/2})$ at $\gamma \ll 1$ and
%as $\ln R$ at $\gamma \le 1$, where $R$ is the upper radius cutoff.

\paragraph{Multiscale expansion.} We shall consider the kink -- vortex
interaction in the limit $\gamma \gg 1$. This rather technical choice is
motivated mainly by simplification of the analysis while retaining results
qualitatively valid olso in the case $\gamma=O(1)$. Let us note, however,
that a rather large
value $2\gamma+1=7$ was reported in \cite{bz}. Under these conditions, a vortex
has a hollow core of unit size, whereas
% a kink
%is thin and rigid.
the excess energy of the kink is concentrated in a
narrow $O(\gamma^{-1/2})$ layer, and $\sigma =O(\gamma^{1/2}) \gg 1$.
Suppose
that the kink separates a positively polarized
fluid ``above" from a negatively polarized fluid ``below", and that a
vortex of negative unit charge is placed in the upper fluid at a distance
$a$ from the kink. We shall
%characterize the kink by its line tension
%$\sigma(\gamma) \gg 1$, and
look for a solution in form of an expansion in
powers of $\sigma^{-1}$. In the zero approximation, the kink is a straight
line, taken as the axis $y=0$. The vortex motion would be determined in
this approximation by its interaction with its image, resulting in
translation parallel to the straight-line kink
%.
%The translation speed $c$
%is obtained exactly in the same way as for a standard case of ideal fluid
%flow past a rigid wall.
%For a vortex of negative unit charge at the
%distance $a$ from the kink, we have
with the velocity $c=-1/(2a)$.

In the next approximation, the kink deformation under the action of the
pressure field due to the moving vortex can be computed. We shall see that
the vortex induces on a kink a capillary wave that takes away the energy of
the vortex leading eventually to annihilation of the vortex at the kink.
%
%The general method is rather similar to a common treatment of acoustic
%waves generated by vortex motion \cite{kambe}.
% Different approximations,
%corresponding to different relative scaling of spacial variables and time,
%are used in an inner ``convective'' region where the fluid motion is
%quasistationary in an appropriately defined moving frame, and an outer
%``acoustic'' region where wave motion prevails. The outer limit of the
%inner solution is taken as the inner limit of the outer solution, i.e. as
%the source of outgoing sound waves. The outer limit of the outer solution
%yields the radiative energy loss that can be used to describe adiabatic
%evolution of parameters of the convective region, e.g. of the distance
%between interacting vortices.
%
%In our case, the superfluid can be assumed incompressible, and the
%principal radiative effect dissipating the energy of the kink -- vortex
%configuration is radiation of {\it capillary} waves.
%The general approach
%distinguishing between the regions of prevailing quasistationary and wavy
%motion remains, however, intact. This approach has been applied to a
The approach that we use has been applied formerly to a
somewhat similar system -- interaction of a vortex with a slightly
deformable surface of an ideal fluid -- by Novikov \cite{nov}. There are
two principal distinctions between our system and that of Ref.~\cite{nov}.
First, instead of one fluid separated by the interface from an inviscid
medium, we have two superfluids with identical physical properties on both
sides of the interface. Second, a {\it capillary} wave is excited in our
case, as
opposed to a {\it gravity} wave in Ref.~\cite{nov}.
%The latter distinction is
%most important, as a qualitative difference in dispersion relationships
%leads, as we shall see, to a reversal in the direction of the propagating
%wave relative to the direction of vortex motion.

Since a thin slightly bent kink
%(with the curvature radius far exceeding
%its thickness)
can be treated as a usual fluid interface possessing a
certain line tension $\sigma$, its small deformation $h(x)=O(\sigma^{-1})$
is determined by two
boundary conditions: the kinematic boundary condition
\begin{equation}
h_t = w^\pm- v^\pm h_{x} + hw^\pm_y,
\label{kinematic}  \end{equation}
and the normal stress balance
\begin{equation}
\sigma h_{xx} = p^{(+)}- p^{(-)} ,
\label{surface}  \end{equation}
Here $x$ and $y$ are coordinates directed, respectively, along and normally
to the unperturbed kink; $v=\phi^\pm_x$ and  $w^\pm=\phi^\pm_y$ are the
respective velocity components in either fluid computed at the kink line,
and $p^{(\pm)}$ are pressures at both sides of the kink that can be
computed using the Bernoulli equation (\ref{bern}).
%Both
%Eqs.~(\ref{surface}) and (\ref{kinematic}) are written assuming the
%deformation to be small, and, indeed, it follows from Eq.~(\ref{surface})
%that $h=O(\sigma^{-1}) $ at distances of $O(a)$ (see below).
The velocities
and their derivatives in Eq.~(\ref{kinematic}) are computed at the
unperturbed interface position $y=0$, and the last term in
Eq.~(\ref{kinematic}) gives a correction to the vertical velocity at the
actual shifted interface.
%Since both fluids are inviscid, the tangential
%stress balance does not hold, and the tangential velocity component can
%suffer a discontinuity at the kink.

\paragraph{Inner solution.}
%In the inner region $x=O(a)$,  the quasistationary
%deformation of the kink can be computed in
In the coordinate frame comoving
with the vortex,
%. Then
the pressure field is stationary in the leading
approximation, and the terms containing the time derivatives in both the
Bernoulli equation (\ref{bern}) and the kinematic boundary condition
(\ref{kinematic}) can be neglected. Using the Bernoulli equation rewritten
in the moving frame yields the pressure in the ``upper'' fluid (at $y>0$)
containing the vortex: \begin{equation}
-p^{(+)} = -c \left( \phi^{(+)}_x + \phi^{(-)}_x\right) +
     \frac{1}{2} \left| \nabla\phi^{(+)}_x + \nabla\phi^{(-)}\right|^2
     = \frac{a^2-x^2}{(a^2+x^2)^2}.
\label{pressure}  \end{equation}
Here $\phi^{(\pm)}$ are phase fields (flow potentials) due to,
respectively, the vortex at $\{0,a\}$ and its image at $\{0,-a\}$. In the
half-plane $y<0$ there is no motion in the zero approximation, so $p^{(-)}
=0$.

Using Eq.~(\ref{pressure}) in Eq.~(\ref{surface}), and taking the highest
elevation $h(0)$ to be equal to zero, readily yields the solution
\begin{equation}
h = - \frac{1}{2\sigma} \,
    \ln \left[1 + \left(\frac{x}{a}\right)^2\right]
\label{surface1}  \end{equation}

This solution
%should describe reasonably well $O(\sigma)$ surface
%deformations
is valid
sufficiently close to the vortex but diverges logarithmically
at $x \to \pm \infty$.
%The kinematic condition plays in the inner region a
%minor role.
%Once the deformation $h(x)$ has been computed, one can use the
%quasistationary version of Eq.~(\ref{kinematic}), rewritten as $\phi^\pm_y=
%\partial_x(h \phi_x)$ as a boundary condition for computing an $O(\sigma)$
%correction to the velocity field.

\paragraph{Outer region.} At large distances, a different approximation
should be used, taking into account non-stationary effects related to
excitation of capillary waves propagating along the kink. At a distance $l
\gg \sigma \gg a$, the time derivative term in Eq.~(\ref{kinematic})
balances the $O(l^{-1})$ linear term if the characteristic time scale is of
$O(l/\sigma)$. The nonlinear terms in Eq.~(\ref{kinematic}) are of
$O(\sigma^{-1}l^{-2})$ at these distances, and can be neglected. Thus, the
correction to the velocity field can be obtained at large distances by
solving the Laplace equation for the $O(\sigma)$ correction $\widetilde
\phi^\pm$ to the flow potential in both fluids, $\nabla^2
\widetilde\phi^\pm =0$ with the boundary condition $\widetilde \phi^\pm_y =
h_t$ at $y=0$.

%The solution of this problem should be obtained in
In the ``laboratory''
%rather than moving
frame
%. We are looking for a solution in the form of a
%Fourier series, and
we define
\begin{equation}
\hat\phi^\pm_k (y,t) = \int_{-\infty}^{\infty}
\widetilde \phi^\pm (x,y,t) e^{-ikx} dx, \;\;
\hat h_k (y,t) = \int_{-\infty}^{\infty} h (x,y,t) e^{-ikx} dx .
\label{fourier} \end{equation}
The solution is
\begin{equation}
\hat\phi^\pm_k (y,t) = \mp |k|^{-1} \hat h_t e^{-|k|y}
\label{sfourier} \end{equation}

The nonlinear term in Eq.~(\ref{bern}) can be also neglected under these
conditions, so that the Fourier components $\hat p^\pm_k$ of the additional
pressure in both fluids are immediately computed as

\begin{equation}
\hat p^\pm_k (y,t) = \pm |k|^{-1} \hat h_{tt} e^{-|k|y}
\label{pfourier} \end{equation}

The solution is closed using the Fourier transform of the normal stress
boundary condition (\ref{surface})
\begin{equation}
-\sigma k^2\hat h= 2 |k|^{-1} \hat h_{tt} + \hat p^{(0)}e^{-ikct},
\;\;\;
\hat p^{(0)} = \frac{|k|}{4\pi} e^{-|k|a} .
\label{hfourier} \end{equation}
where $\hat p^{(0)}$ is the Fourier transform of the zero-order pressure in the
upper fluid given by Eq.~(\ref{pressure})
.
%:
%\begin{equation}
%\hat p^{(0)} = - \int_{-\infty}^{\infty} \frac{a^2-x^2}{(a^2+x^2)^2}
% e^{ikx} dx = \frac{|k|}{4\pi} e^{-|k|a} .
%\label{fpressure}  \end{equation}
%The explicit time dependence in the last term in Eq.~(\ref{hfourier})
%appears after Eq.~(\ref{pressure}) obtained in the comoving coordinates is
%translated back to the laboratory frame.

It is convenient to rewrite Eq.~(\ref{hfourier}) in the form
\begin{equation}
\hat h_{tt} + \omega^2\hat h= f(k) e^{-ikct},
\label{hf} \end{equation}
where $f(k)=-(k^2/8\pi)e^{-|k|a},\; \omega^2 = \frac{1}{2}\sigma |k|^3$.
The forcing term in the right-hand side is {\em resonant} at $\omega = kc$.
This resonance gives the most important contribution to the surface
deformation that would not decay at large distances and long times. In
order to obtain the asymptotic form of the resonant wave, we introduce,
following Novikov \cite{nov}, a modified forcing $f(k)e^{(-ikc +
\epsilon)t}$, and take subsequently the limit $\epsilon \to 0$. Solving
Eq.~(\ref{hf}) with the modified forcing yields
\begin{equation}
h = \frac{e^{\epsilon t}}{2\pi} \int_{-\infty}^{\infty}
\frac{f(k) e^{ik(x-ct)} dk}{\omega^2 - (kc+i\epsilon)^2}.
\label{hsolve}  \end{equation}
The integral is computed by closing a contour in the complex
$k$-plane.
%This should be done by drawing a large half-circle lying in the
%upper or lower half-plane, respectively, at $x>0$ and $x<0$. In both cases,
%the contour should go around the cut along the imaginary axis.
%
%The integrand in Eq.~(\ref{hsolve}) has four simple poles, two of them,
%$k=\pm1 - 2i\epsilon$, lying in the lower, and
%two, $k=\pm[i\epsilon +2^{-1/2}(1-i)\epsilon^{3/2}]$, in the upper half-plane.
%One can show that the integral along the cut decays as $|x|^{-1}$ at $|x|
%\to \infty$ in both positive and negative $x$. Also, the contribution of
%poles in the upper half-plane vanishes in the limit $\epsilon \to 0$, and
%only the contribution of poles in the lower half-plane both remains finite
%in this limit and does not decay at large distances. Thus, the asymptotic
%solution is
The asymptotic solution as $|x|\to \infty$ is
\begin{equation}
h = \left\{ \begin{array}{lll} \frac{1}{\sigma}
   \sin \frac{x-ct}{2\sigma a^2} & \mbox{at} & x<0 \\
    0 & \mbox{at} & x>0 \end{array} \right. .
\label{hfin}  \end{equation}

Obviously, the wave number of the capillary wave generated by the vortex motion
$K=1/(2a^2\sigma)$ satisfies the resonance condition $\omega(K)=|c|K$.
It is remarkable that, in contrast to the case of gravity waves considered by
Novikov \cite{nov}, the wave runs {\em ahead\/} of the vortex. This
circumstance is connected with the fact that the group velocity of
capillary waves exceeds
their phase velocity.

%The generation of the capillary wave on the phase boundary
%diminishes the
%energy of the vortex/image pair
%and leads to their eventual annihilation.
%In order to calculate the energy losses of the vortex/image pair, we find the
The energy flux connected with the wave (\ref{hfin}) that
diminishes the
energy of the vortex/image pair
%using the formula
is \cite{whi}
\begin{equation}
F=C(K)E_w,
\label{flux}  \end{equation}
where $E_w$ is the energy density and $C(K)=3/4a$ is the group velocity of
waves.
The full (surface) energy density is the sum of the deformation energy $E_1$
of the kink and the kinetic energy of both fluids $E_2$.
%The former term is
%\begin{equation}
%E_1=\frac{1}{2}\sigma K^2h_0^2,
%\label{energy1}  \end{equation}
%where $h_0=1/\sigma$ is the amplitude of the wave. The energy of the fluids
%motions
%\begin{equation}
%E_2=\frac{1}{2}\int_{-\infty}^{\infty}v^2 dy=\frac{1}{2}\frac{\omega^2(K)}{K}
%h_0^2=\frac{1}{4}\sigma K^2h_0^2.
%\label{energy2} \end{equation}
% We find that
%\begin{equation}
% E_w=E_1+E_2=\frac{3}{8\sigma^3a^4},
%\label{energyw} \end{equation}
%thus
The calculations show that
\begin{equation}
F=\frac{9}{32\sigma^3a^5}.
\label{fluxw}  \end{equation}
%Taking into account the fact that
Because the energy of the vortex--image pair is
\begin{equation}
E_p=\frac{1}{4\pi}\ln\frac{2a}{\delta},
\label{energy}  \end{equation}
where $\delta$ is the radius of the vortex core, we find that the balance
condition $dE_p/dt=-F$ leads to the following dynamics of the vortex:
\begin{equation}
\frac{da}{dt}=-\frac{9\pi}{8\sigma^3}\frac{1}{a^4}.
\label{dynamics}\end{equation}
Thus, the vortex--image pair tends to annihilate, and the distance between
the vortex and the image is governed by the law
\begin{equation}
a\sim (t_c-t)^{1/5}.
\label{collapse}\end{equation}
Certainly, this law is valid while $a$ is still large compared to the size
of the vortex core.

\paragraph{Conclusion.} We found that the interaction of the vortex in
a circularly polarized component with the phase boundary between phases with
opposite signs of polarization
%does not amount to a trivial drift of the
%vortex along the boundary caused by the interaction with its image. The
%resonant excitation of
generates a specific capillary wave on the boundary. This phenomenon leads to
energy losses by the vortex--image pair that has to terminate in its
eventual annihilation.

\paragraph{Acknowlegement.} A major part of this work was carried out
during the authors' stay at the Institute Henri Poincar\'e, Paris.
%------------

\end{document}